\documentclass[twocolumn,pra,nofootinbib]{revtex4}
\usepackage{amsmath,amsfonts}
\usepackage{graphicx}

\def\be{\begin{equation}}
\def\ee{\end{equation}}
\def\bea{\begin{eqnarray}}
\def\eea{\end{eqnarray}}
\def\nn{\nonumber}

\begin{document}
\title{\bf Ultralight dark photon as a model for early universe dark matter}

\author{V.V. Flambaum$^{1,2}$ and I.B. Samsonov$^{1}$}
\affiliation{$^1$School of Physics, University of New South Wales,
Sydney 2052,  Australia,}
\affiliation{$^2$Johannes
Gutenberg-Universit\"at Mainz, 55099 Mainz, Germany}

\begin{abstract}

Dark photon is a massive vector field which interacts only with
the physical photon through the kinetic mixing. This coupling
is assumed to be weak so that the dark photon becomes almost unobservable in processes
with elementary particles, but can serve as a dark
matter particle. We argue that in very early Universe ($z>3000$) this
vector field may have the equation of state of radiation ($w=1/3$)
but later behaves as cold dark matter ($w=0$). This may slightly change
the expansion rate of the Universe at early time and reduce the
value of the sound horizon of baryon acoustic oscillations (standard
ruler). As a result, in this model the value of the Hubble constant appears to be
larger than that in the standard $\Lambda$CDM model.
In particular, it is sufficient to have the dark photon mass of order $m\sim
10^{-27}-10^{-25}$ eV to fit the value of the Hubble constant to $H_0 = 73$ km~s$^{-1}$Mpc$^{-1}$ thus resolving the Hubble tension.

\end{abstract}

\maketitle

\section{Introduction}

The $\Lambda$CDM cosmological model, in spite of its simplicity, is very
successful in describing the Universe expansion history. Recent
measurements of the cosmic microwave background anisotropy \cite{Plank}
specify the parameters of the $\Lambda$CDM model with a very high
precision. In particular, the inferred value of the Hubble
constant is $H_0=67.4\pm 0.5$ km~s$^{-1}$Mpc$^{-1}$, which,
however, disagrees with the results from
supernovae \cite{NS-Hubble,NS-Hubble1} and lensing time delays
\cite{Bonvin:2016crt,Birrer:2018vtm}. The latter experiments measure a greater
value for $H_0$ of about $73$ km~s$^{-1}$Mpc$^{-1}$. This discrepancy, known
also as the ``Hubble tension,'' may indicate that the standard
$\Lambda$CDM model is incomplete and requires some modifications.

As discussed in \cite{NS-Hubble}, one of the possible resolutions of
the Hubble tension is based on the assumption that at the early stage of the Universe's
expansion (during the radiation dominated epoch) there might be extra
relativistic particle species which contributed to the radiation
density. Such particles could be sterile neutrinos or
any other light particles not accounted within the Standard Model of
elementary particles. With these extra contributions to the
radiation density, the expansion rate of the Universe would
be larger, and the recombination epoch would start earlier. This
reduces the value of the sound horizon for baryon acoustic
oscillations (BAO) and, thus, dictates a larger value for $H_0$.

Following the above idea, it is natural to assume that in the
early Universe there were extra vector fields, different from the visible
photon field, which might contribute to the radiation density in the
pre-recombination epoch. As they are unobservable now, these vector fields
must interact very weakly with the visible matter and should
have a small mass to describe stable particles. In particular, the
so-called dark photon field which was introduced originally in \cite{DP1,DP2}
as a model of dark matter particles may play this role. However,
we will consider a general massive vector field which is
not necessarily identified with the dark photon.

In this paper we present a solution of equations of motion of the
massive vector field in the radiation dominated expanding Universe
which behaves as radiation before certain time, but then changes to the
cold dark matter state. Using this solution, we propose a modification of the
standard $\Lambda$CDM model in which some fraction of the
dark matter is composed of the massive vector fields. We then show
that it is possible to fit the parameters of this model such that
the value of the sound horizon is reduced by about 6\%, and the
inferred value of the Hubble constant becomes $73$ km~s$^{-1}$Mpc$^{-1}$.

As was demonstrated in recent publications \cite{Poulin:2018cxd,Poulin:2018dzj},
a similar result may be achieved with the use of a scalar field,
which may be identified with the axion-like particles. The model
presented in these papers, however, is based on a modification of
the equation of state of the dark energy at early time while in this paper
we propose to modify the equation of state of dark matter. Indeed,
the equation of state of a ``frozen'' scalar field is equivalent
to the cosmological constant which plays minor role at early
Universe. A useful feature of the vector field model considered in this paper
is that the vector field automatically behaves as radiation in
early Universe even for the simplest model of the massive vector field
minimally coupled to gravity, but after some critical time (which is equal to the
inverse mass of the vector field) it turns to the state of cold dark
matter. Therefore, our model represents an extension of the $\Lambda$CDM
model such that some (small) fraction of the dark matter is
``hot'' during a short period of time in early Universe.

One of the typical issues with vector fields in cosmology is that
any particular vector field solution may create a preferred direction
and break the isotropy of the Universe. The isotropy of the Universe
may be preserved in one of the two ways: (i) with the use of a triplet of
mutually orthogonal vector fields with the same mass and the same
magnitude or (ii) by considering a large number of randomly
oriented vector fields. As was demonstrated within the vector
inflation model \cite{Mukhanov}, the latter model predicts a
small anisotropy of the Universe while the triplet model provides
fully isotropic solution. Although both cases are interesting, in this paper
we will focus only on the triplet model which exactly preserves the isotropy
of the Universe.

One may also question about mechanisms of production of the vector
fields in early Universe. As was shown in \cite{Graham}, ultra-light vector
fields may be generated from inflationary fluctuations in a
sufficient abundance to model the dark matter. Although the
authors of \cite{Graham} considered the vector fields with the mass
$m \gtrsim 10^{-5}$ eV, they did not exclude the scenario that
some fraction of the dark matter might be described by much lighter
vector fields homogeneous in space. Note also that alternative
scenarios of generating relic massive vector fields were proposed
in Refs.\ \cite{1,2,3}.

The rest of this paper is organized as follows. In section
\ref{Sec2} we consider a solution of the massive vector field in
the radiation-dominated expanding Universe and study the equation of
state corresponding to this solution. In section \ref{Sec3} we
use the obtained solution to modify the standard $\Lambda$CDM
model so that at early time of the Universe's expansion there is
an excess of radiation density which slightly changes the
expansion rate and the Hubble
constant. The last section is devoted to a discussion of the
presented model.

\section{Massive vector field in expanding Universe}
\label{Sec2}
We start this section by briefly revisiting the massive vector field model
in the expanding Universe with the focus on the equation of state of this field in the
spatially homogeneous case. We then consider a particular solution of the equation of motion for
this field corresponding to the radiation-dominated epoch.

\subsection{Homogeneous massive vector field in expanding Universe}

Let us consider a homogeneous and isotropic expanding Universe
with the metric $g_{\mu\nu}={\rm diag}(-1,a^2,a^2,a^2)$, where
$a=a(t)$ is the scale factor. The Lagrangian of a massive vector
field $A_\mu$ on this background reads
\be
L= \sqrt{-g}\left(
-\frac14 F_{\mu\nu}F^{\mu\nu}
- \frac{m^2}2
A^\mu A_\mu
\right)\,,
\ee
where $F_{\mu\nu} = \partial_\mu A_\nu - \partial_\nu A_\mu$ and
$g=\det g_{\mu\nu}=-a^6$. The corresponding equation
of motion (Maxwell-Proca equation) reads
\be
-\frac1{\sqrt{-g}}\partial_\mu \sqrt{-g}F^{\mu\nu} + m^2 A^\nu =0\,.
\label{EOM}
\ee
The conserved stress-energy tensor has standard form
\be
T_{\mu\nu} = F_{\mu\alpha}F_\nu{}^\alpha - \frac14 g_{\mu\nu}
F_{\alpha\beta} F^{\alpha\beta}
-\frac{m^2}{2} g_{\mu\nu} A_\alpha A^\alpha + m^2 A_\mu A_\nu\,.
\label{T}
\ee

Let ${\bf A}$ and $\phi$ be 3+1 components of the vector field, $A_\mu = (-\phi, {\bf
A})$, ${\bf A}\equiv (A_i)=(A_1,A_2,A_3)$. In terms of these components, the electric and magnetic
fields may be defined as
\be
{\bf E} = -\dot{\bf A} - \nabla\phi\,,\quad
{\bf B} = \nabla\times{\bf A}\,,
\label{EB}
\ee
where the dot over the field stands for the time derivative, e.g.,
$\dot{\bf A}\equiv \partial_t {\bf A}$.
In the non-covariant form, the equation of motion (\ref{EOM})
reads
\begin{subequations}
\label{non-covEOM}
\bea
\nabla \cdot {\bf E} + m^2 a^2 \phi &=&0\,,\label{11}\\
-\dot {\bf E} - \frac{\dot a}{a}{\bf E} + \frac1{a^2} \nabla\times {\bf B} +
m^2 {\bf A} &=&0\,.
\label{12}
\eea
\end{subequations}

We will look for a homogeneous solution of the equations
(\ref{non-covEOM}), $\partial_i A_\mu=0$. In this case, Eq.\
(\ref{11}) implies $\phi=0$ and ${\bf A}={\bf A}(t)$, while Eq.\
(\ref{12}) reduces to
\be
\ddot{\bf A} + \frac{\dot a}{a} \dot{\bf A} + m^2 {\bf A} =0\,.
\label{Aeq}
\ee
A solution of this equation gives homogeneous electric field with
vanishing magnetic field, ${\bf B}=0$.

The 3+1 components of the stress-energy tensor (\ref{T}) are
\bea
T_{00} &=& \frac1{2a^2} ({\bf E}^2 + m^2 {\bf A}^2)\,,
\label{T00}\\
T_{ij} &=& -E_i E_j + \frac12\delta_{ij} {\bf E}^2
+ m^2 A_i A_j - \frac{m^2}2 \delta_{ij} {\bf A}^2\,,
\label{Tij}
\eea
and $T_{0i}=0$ for the homogeneous vector field subject to
Eq.\ (\ref{Aeq}). Note that the tensor
$T_{ij}$ is non-diagonal since the electric field creates
anisotropy of the Universe.

Since no anisotropy of the Universe is
observed \cite{Plank}, the
stress-energy tensor (\ref{Tij}) should have the diagonal form.
The isotropy of the expanding Universe driven by vector fields
may be naturally achieved in one of the two ways: (i) using a
triplet of mutually orthogonal vector fields with the same mass
and magnitude and (ii) by applying a large number $N$ of randomly
oriented vector fields which have no preferred direction in
average. In the latter case, as was demonstrated in the vector inflation model
\cite{Mukhanov}, the off-diagonal components of the stress-energy
tensor are not exactly vanishing, but are proportional to
$\sqrt{N}$ predicting a small anisotropy of the Universe. Although
this case may be interesting, in this paper we will consider the
fully isotropic vector field model based in the triplet of
orthogonal vector fields.

Let $A^{(a)}_\mu$, $a=1,2,3$, be a triplet of mutually orthogonal
vector fields with the same mass $m$ and
magnitude $|{\bf A}|$. One can prove the following simple identities by
averaging over the species:
$\overline{A_i A_j} \equiv \frac13 \sum_a A^{(a)}_i A^{(a)}_j
=\frac13 \delta_{ij}{\bf A}^2$, $\overline{E_i E_j}
\equiv \frac13 \sum_a E^{(a)}_i E^{(a)}_j=\frac13 \delta_{ij}{\bf
E}^2$. As a result, the tensor (\ref{Tij}) in average acquires the diagonal
form
\be
\overline{T_{ij}} = \frac16\delta_{ij} ({\bf E}^2 - m^2 {\bf A}^2)\,.
\label{Tij1}
\ee

The equations (\ref{T00}) and (\ref{Tij1}) define the energy
density $\rho = T^{00}$ and pressure
$\overline{T_i^j}=p\delta_i^j$ created by the triplet of
massive vector fields,
\bea
\rho &=& \frac1{2a^2} ({\bf E}^2+ m^2 {\bf A}^2)\,,\label{10}\\
p&=& \frac1{6a^2}({\bf E}^2 - m^2 {\bf A}^2)\,.
\eea
The corresponding equation of state reads
\be
w\equiv\frac p\rho = \frac13\frac{{\bf E}^2 - m^2 {\bf A}^2}{{\bf E}^2+ m^2 {\bf
A}^2}\,.
\label{w}
\ee
In the massless case, $m=0$, this parameter corresponds to the
equation of state of radiation, $w=\frac13$, but for non-vanishing
mass it changes in the interval $-\frac13\leq w \leq \frac13$.

Note also that Eq.\ (\ref{w}) resembles the equation of state of
the constant magnetic field solution in massive electrodynamics considered
in \cite{Ryutov:2017oeq}.

\subsection{Massive vector field solution in early Universe}

In the radiation-dominated epoch, the scale factor $a$ is well approximated by
\be
a(t) = (2\sqrt{\Omega_{\rm r}}H_0 t)^{1/2}\,,
\label{arad}
\ee
where $\Omega_{\rm r}$ is the radiation density and $H_0$ is the
Hubble constant.
With this scale factor Eq.\ (\ref{Aeq}) reduces to
\be
\ddot{\bf A} + \frac{1}{2t} \dot{\bf A} + m^2 {\bf A} =0\,.
\label{Aeq1}
\ee
The general solution of this equation reads
\be
{\bf A}(t) = {\bf A}_0 \cdot (mt)^{\frac14}
\left(c_1 J_{\frac14}(mt) + c_2  Y_{\frac14}(mt)\right)\,,
\label{Asol}
\ee
where $J$ and $Y$ are the Bessel functions, ${\bf A}_0$ is a
constant vector, and $c_1$, $c_2$ are the integration constants.%
\footnote{More generally, the solution of Eq.\ (\ref{Aeq1})
may be written as $(mt)^{\frac14}
\left({\bf A}_1 J_{\frac14}(mt) + {\bf A}_2  Y_{\frac14}(mt)\right)$
with two arbitrary constant vectors ${\bf A}_1$ and ${\bf A}_2$.} The
corresponding electric field is found from Eq.\ (\ref{EB}),
\be
{\bf E}(t) = -{\bf A}_0\cdot m(mt)^{\frac14}
\left(c_1 J_{-\frac34}(mt) + c_2 Y_{-\frac34}(mt)\right)\,.
\label{Esol}
\ee
For this solution, the energy density (\ref{10}) reads
\bea
\rho&=&\frac{|{\bf A}_0|^2m^2\sqrt{mt}}{2a^2}
\bigg[\left(c_1 J_{-\frac34}(mt)+c_2 Y_{-\frac34}(mt)\right)^2
\nn\\&&
+\left(c_1 J_{\frac14}(mt)+c_2 Y_{\frac14}(mt)\right)^2\bigg]
\,.
\label{density}
\eea

The energy density (\ref{density}) has the following asymptotics
for $c_1+c_2\ne0$:
\be
\rho\approx |{\bf A}_0|^2\left\{
\begin{array}{l}
\frac{\sqrt{2}m(c_1+c_2)^2}{\Gamma^2(\frac14)}
\frac{1}{a^2 t}\quad\mbox{ for }\quad mt\ll1\\
2m^2(c_1^2+c_2^2)\frac{1}{a^2\sqrt{mt}}
\quad\mbox{ for }\quad mt\gg1\,.
\end{array}
\right.
\ee
Thus, taking into account Eq.\ (\ref{arad}), we conclude that
the energy density of the homogeneous massive vector field
in the early Universe scales as radiation for $t\ll m^{-1}$ and
as cold dark matter for $t\gg m^{-1}$,
\be
\rho\propto\left\{
\begin{array}{l}
a^{-4}\quad\mbox{ for }\quad t\ll m^{-1}\\
a^{-3}\quad\mbox{ for }\quad t\gg m^{-1}\,.
\end{array}
\right.
\ee
Therefore, the equation of state of this field may be approximated
by the step function,
\be
\bar w= \left\{
\begin{array}{ll}
\frac13 \ &\mbox{ for } t< m^{-1} \\
0 &\mbox{ for } t> m^{-1}\,.
\end{array}
\right.
\label{barw}
\ee

The above solution is valid for arbitrary initial conditions
except for $c_1+c_2=0$. In the latter case, for $mt\ll1$ the
energy density has the following asymptotics
$\rho\approx \frac{m^2 \Gamma^2(\frac14)}{\sqrt2\pi^2a^2}$ which
corresponds to $w=-\frac13$. We assume that this case is not
realized in Nature since it requires special initial conditions.

We stress that the massive vector field solution (\ref{Asol}) is
not universal as it applies only in the radiation-dominated epoch.
At later times, e.g., in the matter-dominated epoch, the behavior
of the vector field may change, but its density should be negligible
compared with the matter density (including the dark matter) to
avoid significant changes in the expansion rate of the late Universe.

\section{Implication to the Hubble tension problem}
\label{Sec3}
For completeness of our presentation, we start this section
with a short review of the standard $\Lambda$CDM model. We then
present a modification of this model which may resolve the Hubble
tension problem while keeping all other conclusions of this model
intact.

\subsection{$\Lambda$CDM model and the standard ruler}

The Hubble parameter $H \equiv \frac{\dot a }{a} $ in the $\Lambda$CDM model is given by
\be
H(a)\equiv H_0 E(a) =
H_0 \sqrt{\Omega_{\rm r} a^{-4} + \Omega_{\rm m} a^{-3}+ \Omega_\Lambda}\,,
\label{Ha}
\ee
where $H_0$ is the Hubble constant and $\Omega_{\rm r}$, $\Omega_{\rm m}$, and
$\Omega_\Lambda$ are density parameters of the radiation,
matter and cosmological constant, respectively. Here we consider
the spatially flat Universe, for simplicity. Note that
$\Omega_{\rm m}$ takes into account both baryonic and cold dark
matter, $\Omega_{\rm m}=\Omega_{\rm b}+\Omega_{\rm c}$, while the radiation
density is the sum of photon and neutrino contributions, $\Omega_{\rm r}=
\Omega_\gamma+\Omega_\nu$. The values of these parameters can be
taken from the Particle Data Group review \cite{PDG}:
$H_0 = 67.8$ km s$^{-1}$ Mpc$^{-1}$, $\Omega_{\rm b} = 0.048$,
$\Omega_{\rm c}=0.258$, $\Omega_\gamma = 5.37\times
10^{-5}$, $\Omega_\nu = 3.66\times 10^{-5}$, $\Omega_\Lambda =
0.692$. Note that the density parameter $\Omega_\nu$ corresponds
to relativistic neutrinos since we are going to apply Eq.\
(\ref{Ha}) at early times of the Universe expansion.

The early Universe may be considered as a hot plasma composed mainly of
baryons, electrons and photons. Density fluctuations in this
plasma are known to propagate in the form of sound waves with the
speed \cite{Ma:1995ey}
\be
c_s(a) = \frac{c}{\sqrt{3\left(1+\frac{3a\Omega_{\rm b}
}{4\Omega_\gamma}\right)}}\,,
\ee
where $c$ is the speed of light. Such sound waves propagate
until the decoupling of photons that
occurs ar redshift $z_*\approx 1090$. As a result, initial
perturbations generate spherical waves in
the primordial plasma with the comoving sound horizon, which is
also known as the ``standard ruler,''
\be
r_s = \frac1{H_0} \int_0^{a_*} \frac{c_s(a)da}{a^2 E(a)}\,,
\label{rs}
\ee
where $a_* = (z_*+1)^{-1}$. Direct numerical evaluation of this
integral with the above given parameters of the $\Lambda$CDM model yields
$r_s\approx 145$ Mpc. This estimate is in a good agreement with
the WMAP CMB observations \cite{WMAP} which constrain this parameter
as $r_s = 146.8\pm 1.8$ Mpc.

The BAO data allows one to consider the sound horizon as a standard
comoving ruler, whose length is independent of redshift and
orientation, but is not necessarily correlated with CMB data which
fix the parameters of the $\Lambda$CDM model and the Hubble
constant. As a result, BAO data may be used to impose the
constraint on the product $(r_s\cdot H_0)$ \cite{Aubourg}:
\be
\frac c{r_sH_0} = 29.63^{+0.48}_{-0.45}\,,
\label{H0constr}
\ee
where $c$ is the speed of light.

In what follows, we will consider an extension of
the $\Lambda$CDM model which predicts a reduced value of the sound
horizon $r_s$ and a larger value for $H_0$
according to the constraint (\ref{H0constr}). We will show that it
is possible to fit the parameters of this model such that the
predicted value of the Hubble constant agrees with the
local measurements of this
constant from cephidae and supernovae observations: $H_0 =  73.24 \pm 1.74  \mbox{ km
s}^{-1}\mbox{Mpc}^{-1}$ \cite{NS-Hubble,NS-Hubble1}.

\subsection{Extension of the $\Lambda$CDM model}

We will consider an extension of the $\Lambda$CDM model in which
some fraction $\Omega_A$ of the dark matter is described by the massive vector
field ${\bf A}$ subject to the equation (\ref{Aeq1}). We assume
that this field was produced non-thermally and was present as a relic
after the inflation epoch.\footnote{For non-gravitational mechanisms of production
of vector fields in early Universe see, e.g., \cite{Graham}.}
Assuming also that this field was produced homogeneously with
the equation of state $w=\frac13$ at the beginning of
radiation-dominated epoch ($t=0$), its equation of state at later
times may be approximately described by the function (\ref{barw}).
This field contributes to the expansion rate of the Universe
through the following modification of the Hubble parameter (\ref{Ha}):
\be
E(a)=\sqrt{\frac{\Omega_{\rm r}}{ a^{4}}+\frac{\Omega_{\rm m}-\Omega_A}{a^{3}}
 +\frac{\Omega_A}{ a^{3(w+1)}}
+\Omega_\Lambda}\,.
\label{E1}
\ee
Here we added the term with $\Omega_A$ which describes the massive
vector field with the equation of state $w$ and reduced the matter
density by the same amount in order to keep the balance $\Omega_{\rm r}+
(\Omega_{\rm m}-\Omega_A)+\Omega_A+\Omega_\Lambda=1$.

In this model, we have two free parameters: the energy density
$\Omega_A$ and the mass of the vector field $m$ which enters the
equation of state as in Eq.\ (\ref{density}). Both these
parameters are assumed to be small enough to provide minor violations
from the standard $\Lambda$CDM model.

In the particular case, when the equation of state is approximated
by Eq.\ (\ref{barw}), the Hubble function (\ref{E1}) may be
written as
\be
E(a)=\left\{
\begin{array}{l}
\sqrt{\frac{\Omega_{\rm r}+\Omega_A}{ a^{4}}+\frac{\Omega_{\rm
m}-\Omega_A}{a^{3}}
+\Omega_\Lambda}\,, \quad mt<1\\
\sqrt{\Omega_{\rm r} a^{-4}+\Omega_{\rm m}a^{-3}+\Omega_\Lambda}\,,\quad
mt>1\,.
\end{array}\right.
\label{E2}
\ee
This function shows that at early times ($t<m^{-1}$) the massive vector field
behaves as the radiation and gives additional contributions to
the radiation density, but it quickly changes its equation of
state such that after the critical time $t=m^{-1}$ it behaves as cold dark matter.
The extra contributions to the radiation density lead to the
increase of the expansion rate of the Universe at early times and,
thus, to changes of our estimates of the Hubble constant obtained
within the extended $\Lambda$CDM model.

In the modified $\Lambda$CDM model with the Hubble function (\ref{E2}),
the formula for the sound horizon (\ref{rs}) has the form
\bea
r_s &=& \frac{1}{H_0} \int_0^{a_1}\frac{c_s(a)da}{\sqrt{\Omega_{\rm r}+\Omega_A
+(\Omega_{\rm m}-\Omega_A)a+\Omega_\Lambda a^2}}\nonumber\\&&
+\frac{1}{H_0} \int_{a_1}^{a_*}\frac{c_s(a)da}{\sqrt{\Omega_{\rm
r}+\Omega_{\rm m} a+\Omega_\Lambda a^2}}\,,
\label{r-mod}
\eea
where $a_1=\left( \frac{2\sqrt{\Omega_{\rm r}}H_0}{m} \right)^{\frac12}$ and $a_* =
1/1091$.

To perform numerical estimates, we have to make some
assumptions about the value of the density parameter $\Omega_A$.
This parameter should not be much smaller than the density of
radiation; otherwise the effect of the massive vector field would
be negligible in the evolution of the Universe, $0.1 \Omega_{\rm r} \lesssim
\Omega_A$. This parameter is also bounded from above by the
requirement that the massive vector field should not change the
expansion rate at the matter dominated epoch driven mainly by the
dark matter, $\Omega_A \ll \Omega_{\rm c}$. More precisely, we
require that $\Omega_A$ should be smaller than the uncertainty in
the measurements of the dark matter density which is
$\Omega_{\rm c}=0.258\pm 0.00435$ \cite{PDG}. Therefore, we will
consider the energy density parameter of the massive vector field in the
interval
\be
9.0\times 10^{-5} < \Omega_A < 4.4\times 10^{-3}\,.
\ee

The mass parameter $m$ may be limited by the applicability of the
considered solution (\ref{Asol}). Since this solution holds only
in the radiation-dominated epoch, we should apply it only for
redshifts $z>3600$. When the Universe passes to the
matter-dominated epoch, the equation of state of this solution (\ref{barw})
should change to the matter type. This constrains the mass of the
vector field through the equation (\ref{arad}), $m>2\cdot
3600^2\sqrt{\Omega_{\rm r}}H_0$, or
\be
m>3.6\times 10^{-28}\mbox{ eV}.
\ee

\begin{figure}[htb]
\begin{center}
\includegraphics[width=8.5cm]{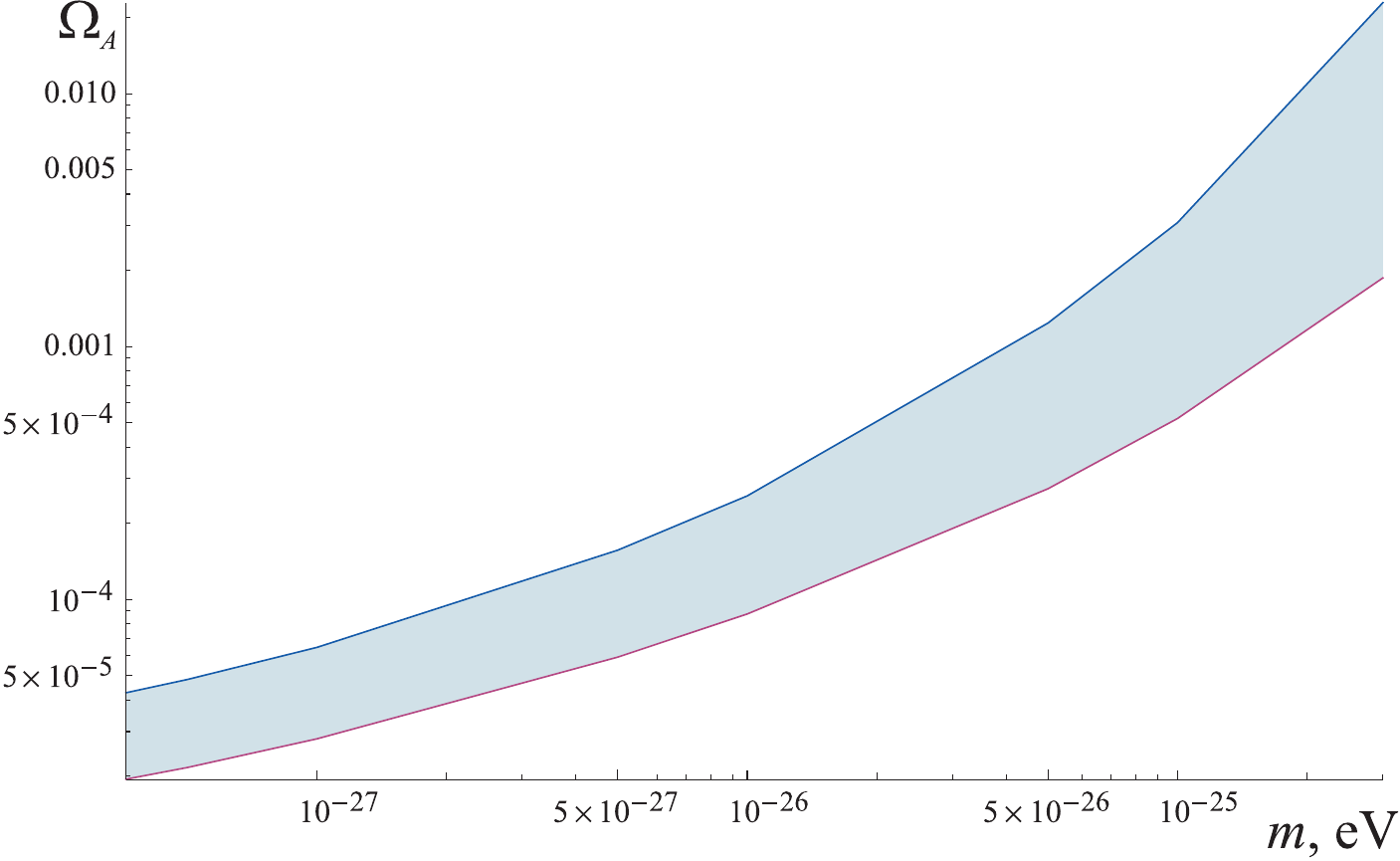}
\caption{Allowed region for the parameters of mass $m$ and density $\Omega_A$
in the massive vector field model.
\label{fig1}}
\end{center}
\end{figure}

We can now use the relation (\ref{H0constr}) to find the value of
the Hubble constant $H_0$ via the known value for the sound
horizon $r_s$ obtained within the modified $\Lambda$CDM model.
We find that the Hubble constant gets the value $H_0=
73$~km~s$^{-1}$~Mpc$^{-1}$ when the parameters $m$ and
$\Omega_A$ lie within the shaded region on the graph in Fig.\
\ref{fig1}.

\section{Conclusions}

In this paper, we propose a modification of the standard
$\Lambda$CDM model in which a (small) fraction of the dark
matter is described by a massive vector field. We consider the
general homogeneous solution of the equation of motion of this
field in expanding Universe during the radiation dominated epoch.
To preserve the isotropy of the Universe, we consider a
triplet of mutually orthogonal fields with the same mass and
magnitude. In this case, we show that the energy density of this solution
behaves as radiation ($w=\frac13$) during the
time $t<m^{-1}$, but for $t>m^{-1}$ it changes to the
cold dark matter ($\bar w=0$).

We consider an implication of the massive-vector-field extended $\Lambda$CDM model
to resolve the Hubble tension. As was argued in \cite{NS-Hubble}, this
tension may be naturally resolved by increasing the number of
relativistic particle species during the radiation-dominated
epoch. Since the massive vector field behaves as radiation at
early times, it may be naturally used to resolve this issue. We
demonstrate that when the mass of the vector field is in the
interval $10^{-27}-10^{-25}$ eV, and the energy density is of order
$\Omega_A\sim 10^{-5}-10^{-2}$, the massive vector field slightly enhances
the expansion rate of the Universe and reduces the sound horizon
of baryon acoustic oscillation by about 6\%. As a result, the
inferred value of the Hubble constant appears to be 73~km~s$^{-1}$Mpc$^{-1}$, which is in agreement with the results from
supernovae \cite{NS-Hubble,NS-Hubble1} and lensing time delays
\cite{Bonvin:2016crt,Birrer:2018vtm}.

We stress that a similar result may be obtained with the use of a
scalar axion-like field \cite{Poulin:2018cxd,Poulin:2018dzj},
which plays the role of dark energy with very specific equation of
state. The main feature of the vector field model is that it
corresponds to the modification of dark matter rather than the
dark energy and automatically behaves as radiation at early times.

The massive vector field considered in this paper may be naturally
identified with the so-called dark photon field introduced
originally in \cite{DP1,DP2} and used in many subsequent
publications as a model for dark matter. There are strict
constraints on the parameters of this model (mass and coupling constant)
from different experiments, see, e.g.,
\cite{DPrev1,DPrev2,DPrev3,DPrev4,DPmy}. In this paper, however,
we consider the ultralight massive vector field which is not
necessary identified with the dark photon since we do not
specify its interaction with visible matter. This interaction is
assumed to be very weak to prevent the thermalization of the
vector field in the early Universe plasma.

\subsection*{Acknowledgments}

This work is supported by the Australian Research Council
Grant No. DP150101405 and by a Gutenberg Fellowship.

\end{document}